\documentclass[reprint,prb,superscriptaddress,showpacs]{revtex4-2}%
\usepackage{amssymb}
\usepackage{xcolor}
\usepackage{color}
\usepackage{graphicx}
\usepackage{epsfig}
\usepackage{float}
\usepackage{soul}
\sethlcolor{green}
\usepackage{amsmath}
\usepackage{amsfonts}
\usepackage{amssymb}
\usepackage{framed}
\usepackage{float}%
\DeclareMathOperator{\sininv}{\sin^{-1}}
\usepackage{amsfonts}
\providecommand{\U}[1]{\protect\rule{.1in}{.1in}}
\makeatletter
\renewcommand*{\fnum@figure}{{\normalfont\bfseries \figurename~\thefigure}}
\renewcommand*{\@caption@fignum@sep}{\textbf{ : }}
\makeatother
\usepackage{hyperref}
\hypersetup{
    colorlinks=true,
    linkcolor=green,
    citecolor=blue,
    filecolor=magenta,      
    urlcolor=cyan,
}

\newcommand {\NF}{N_{\rm F}}
\newcommand {\EF}{\epsilon_F}

\newcommand {\bk}{\mathbf k}
\newcommand {\bq}{\mathbf q}
\newcommand {\bkp}{\mathbf {k'}}

\newcommand {\ek}{\epsilon_{\bk}}

\newcommand {\oql}{\omega_{\bq \nu}}

\newcommand {\oj}{\omega_j}
\newcommand {\ojp}{\omega_{j'}}
\newcommand {\ojjp}{\omega_{j-j'}}

\newcommand {\afkko}{\alpha^{2} F_{\rm ep}(\bk,\bkp,\omega)}

\newcommand {\afo}{\alpha^{2} F(\omega)}

\newcommand {\gkkl}{g^{\nu}_{\bk, \bkp}}

\newcommand {\deltakp}{\delta(\epsilon_{\bkp}-\EF)}

\begin{document}

\title{Electron-phonon coupling and spin fluctuations in the Ising superconductor NbSe$_{2}$}
\author{S. Das}
\affiliation{Department of Physics and Astronomy, George Mason University, Fairfax, VA 22030}
\affiliation{Quantum Science and Engineering Center, George Mason University, Fairfax, VA 22030}
\author{H. Paudyal}
\affiliation{Department of Physics, Applied Physics, and Astronomy, Binghamton University-SUNY, Binghamton, New York 13902, USA}
\author{E. R. Margine}
\affiliation{Department of Physics, Applied Physics, and Astronomy, Binghamton University-SUNY, Binghamton, New York 13902, USA}
\author{D. F. Agterberg}
\affiliation{Department of Physics, University of Wisconsin, Milwaukee, Wisconsin 53201, USA}
\author{I. I. Mazin}
\affiliation{Department of Physics and Astronomy, George Mason University, Fairfax, VA 22030}
\affiliation{Quantum Science and Engineering Center, George Mason University, Fairfax, VA 22030}

\begin{abstract}

{Ising superconductivity, observed experimentally in NbSe$_{2}$ and similar materials, has generated tremendous interest. Recently, attention was called to the possible role that spin fluctuations (SF) play in this phenomenon, in addition to the dominant electron-phonon coupling (EPC); the possibility of a predominantly-triplet state was discussed and led to a conjecture of viable singlet-triplet Leggett oscillations. However, these hypotheses have not been put to a quantitative test. In this paper, we report first principle calculations of the EPC and also estimate coupling with SF, including full momentum dependence. We find that: (1) EPC is strongly anisotropic, largely coming from the K-K' scattering, and therefore excludes triplet symmetry even as an excited state; (2) superconductivity is substantially weakened by SF, but anisotropy remains as above; and, (3) we do find the possibility of a Leggett mode, not in a singlet-triplet but in an $s_{++}$ -- $s_\pm$ channel.
}

\end{abstract}
\maketitle

\section{Introduction} 

Revolutionary progress in the growth and exfoliation of single atomic layers over the last two decades has led to a new era of  scientific discoveries and technological innovation. Following graphene, transition metal dichalcogenides (TMDs) have taken the spotlight, as treasure trove for a plethora of novel quantum phenomena. One of the significant discoveries in recent years was the phenomenon of the so-called Ising superconductivity, driven by spin-orbit (SO) coupling combined with absence of the inversion symmetry~\cite{ising, ising2,Mak,Khodas, Khodas1, Daniel, proximity, TaS2-NbSe2, MoS2, Sergio}. Proximity effects and interfaces of Ising superconductors with other  monolayer TMDs, such as doped TaS$_{2}$ and TaSe$_{2}$~\cite{TaS2-NbSe2,Makarxiv}, or with two-dimensional (2D) magnetic layered materials, such as CrI$_{3}$ \cite{CrI3mag1,CrI3mag2} and VI$_{3}$~\cite{VI3}, could lead to interesting device applications for quantum information storage and spintronics.

The combination of broken Kramer's degeneracy due to the lack of inversion symmetry and SO coupling in monolayers of 2H-NbSe$_{2}$ leads to splitting of the electronic bands 
near the
$K$ point, and its corresponding inversion counterpart, $K^{\prime}=-K$, in the Brillouin zone (BZ). The magnitude of this splitting due to spin-orbit effects is considerably larger than the superconducting order parameter~\cite{Darshana, Makarxiv}. Because of this splitting, the formally s-wave singlet superconducting state well known in the bulk  NbSe$_{2}$,  splits into two mixed states: singlet (S) and triplet (T) states combine to form an S + T state on one SO partner and an S - T state on the other. The same is true about the inversion-related partners, e.g., the outer Fermi contours around $K$ and $K'$ \cite{Darshana}. The emerging  phenomenon was duly dubbed \textquotedblleft Ising superconductivity\textquotedblright (IS). While in most experimental probes the two IS partners combine to form a (nearly) pure S state, the incipient triplet component manifests itself in many notable ways, most famously in the formally infinite thermodynamic critical field along the $ab$ layer plane.

Recent first principles calculations, combined with some limited experimental data, strongly suggest that bulk NbSe$_{2}$ is close to a magnetic instability, and the undistorted monolayers are even closer~\cite{Darshana,divilov2020interplay,Das2021} (and also likely for similar TMD superconductors). This fact led to speculations that triplet pairing, even if not a leading instability, may play an important role in Ising superconductivity in  NbSe$_{2}$ ~\cite{Darshana}. Recent observation of a low-temperature tunneling mode in  NbSe$_{2}$ monolayers was tentatively interpreted as a singlet-triplet Leggett mode~\cite{Ugeda2}.

Recently, we investigated the full momentum-dependent spin susceptibility~\cite{Ronald} in NbSe$_{2}$ monolayers~\cite{Das2021}, and found that it is rather strongly peaked at a particular wave vector, close to ${\bf q}=(0.2,0)$ in the 2D Brillouin zone. At the same time, experimental and density-functional theory (DFT) calculations of charge density waves~\cite{bulkCDW,bulkCDW2, Calandra, Kvashnin, Margine1} and superconductivity~\cite{Khodas, Sergio, Margine1, Ugeda2, layered-NbSe2} for some bulk~\cite{Darshana,bulkNbSe2, bulkCDW, Sanna} and 2D TMDs~\cite{Darshana, Khestanova, monoCDW, Zheng2019, Giustino2022} have been reported. A subsequent first-principles study claimed~\cite{Darshana} that density functional calculations overestimate the superconducting transition temperature in monolayer NbSe$_{2}$. Together with the indications of strong spin fluctuations (SF) in this class of materials, it strongly suggests that a proper quantitative analysis of the pairing state in NbSe$_{2}$, and likely in other Ising superconductors, is not possible without the simultaneous accounting of the anisotropic electron-phonon coupling (EPC) and SF-induced interaction.

In this paper, we present such an analysis and find several expected and some rather unexpected results. First, in agreement with existing calculations of bulk and 2D TMDs, the standard DFT calculations of EPC strongly overestimate the transition temperature in monolayer NbSe$_2$ (far beyond typical inaccuracies of the method). Second, including on the same footing SF-induced interaction (using the previously calculated SF spectrum~\cite{Das2021}) brings the calculations in agreement with experiment (including a proper frequency cutoff for SF is essential). Third, the calculated EPC is exceptionally anisotropic, with the lion's share of the coupling coming from the same-spin $K-K'$ scattering. The calculated gap distribution, formally speaking, should be visible in tunneling experiments, and it has not been observed so far. We discuss possible reasons for why the small gap on the $\Gamma$ Fermi surface pocket has so far eluded detection.

\section{Background Landscape}

\subsection{Tunneling}

Tunneling experiments are an indispensable tool for the discernment of the quantitative as well as qualitative nature of superconducting order parameter in unconventional superconductivity~\cite{Khodas}. In Ref.~[\onlinecite{Ugeda2}] it was pointed out that the different character of the dominant Nb orbitals on the $\Gamma$ and $K$ Fermi surface pockets
suggests that their tunneling probability through vacuum or insulating barrier should be different. The fact that the calculated superconducting gap is rather different at the two sets of pockets suggest that this issue deserves a closer look. 

One possible explanation for the lack of observation of a smaller gap is that, due to impurity scattering, the gap averages to one uniform value. We do not find this likely.
Indeed, the observed $2\Delta/T_c$ ratio is 
noticeably larger than the weak-coupling value of 3.54, and our calculations 
are far from the strong coupling regime where such an enhancement would be
possible. Rather, our larger ($K$) gap agrees consistently with the experiment.
This calls into question, why the second, smaller gap is not seen in the experiment?
We do not have an answer yet, but we can add to the body of known facts, our calculations
of the partial character of Se $p_z$ at the Fermi level. Indeed, in STM
experiments it is rather clear that the main signal comes from Se atoms, and this orbital is the most extended along the out-of-plane direction, so it is expected to dominate the STM spectra.  We show this character as the faux map in Fig.~\ref{fig2}. 

Interestingly, while on average the  $\Gamma$ pocket has a larger content of this character, there are hot spots along the $K-M$ direction that are 
expected to have the largest tunneling probability; taking the calculated value of the superconducting gap at this point yields a rather good agreement with the experiment. On the other hand, while the difference between the tunneling current from $p_z$ orbitals is exponentially higher than that from the $p_{x,y}$ ones, the dependence on the $p_z$ weight is just linear, so, in principle, one would expect to see subgap features corresponding, first of all, to the $\Gamma$ pocket gap approximately twice smaller than the maximal gap.
\vskip 0.2 in
In order to address the nature of superconducting gap, scanning tunneling measurements were performed and reported on few-layer NbSe$_{2}$~\cite{Khestanova}. The superconducting gap as well as the critical temperature ($T_{\rm c}$) have been found to decrease with the number of layers. In particular, the gap values measured at 0.3~K exhibited a reduction by more than a factor of 2 from 1.3~meV in the bulk to 0.6~meV in the bilayer. Unfortunately, no tunneling current was detectable in the monolayer devices, most likely due to the difficulty of obtaining a clean NbSe$_2$-hBN interface. The decrease in the $T_{\rm c}$ has been found to be well described by a linear dependence with the inverse thickness, with the temperature dropping from 7.0~K in bulk to 4.7~-~4.8~K and 2.0~-~2.5~K in bilayer and monolayer, respectively. This drastic decrease in both the measured superconducting gap and critical temperature  has been assigned to the surface energy contribution imposed by the boundary condition upon the electronic wave function. Further, it has been conjectured that while for up to 5 layers or higher, the gap is considerably anisotropic, the anisotropy disappears and the gap obeys the isotropic Bardeen Cooper-Schrieffer (BCS) gap equations for the bilayer~\cite{Khestanova}. The hypothesis that the incommensurate charge density wave is enhanced by the simultaneous existence of superconductivity in monolayer NbSe$_{2}$ has also been proposed.~\cite{layered-NbSe2}

\subsection{Experimental results vs magnetic and electron-phonon coupling calculations}

Superconductivity in bulk NbSe$_2$ has been studied extensively both experimentally and theoretically,  and the superconducting transition temperature $T_{\rm c}$ has been experimentally identified as $\sim$ 7~K ~\cite{Foner}. Compared to bulk, $T_{\rm c}$ of monolayer NbSe$_2$ is about half, up to $\sim$ 3.5~K in best samples (it is often as low as 1~K)~\cite{TaS2-NbSe2, Mak}. It was argued that that is due to the pair-breaking effect of magnetic moments associated with Se vacancies~\cite{proximity}. 

State-of-the-art first-principles calculations that usually deliver accurate outcomes for superconductors where the pairing is entirely due to EPC overestimate the $T_{\rm c}$ in bulk NbSe$_{2}$~\cite{bulkCDW} and isostructural NbS$_{2}$~\cite{Margine1}. In the latter case, calculations using  Eliashberg theory yield a $T_{\rm c}$ and a zero-temperature gap a factor of $\sim$ 3 and 4 larger than experiment, respectively~\cite{Margine1}. 
At the same time, the experimentally measured spin susceptibility, $\chi_{s}$, in bulk NbSe$_{2}$ was reported to be $\chi_s\sim 3\times10^{-4}$ emu/mole {~\cite{exptchi}}, which significantly exceeds the bare bulk Pauli susceptibility $\chi_{0}\sim 0.87 \times 10^{-4}$ emu/mole. DFT calculations render $\chi_s=4.2\times10^{-4}$ emu/mole {~\cite{Das2021, Darshana}}, 40\% larger that in the experiment -- a common overestimation in itinerant systems, indicating that SF are strong in the system.

Recently, we have calculated the static $\bf q$-dependent DFT susceptibility in NbSe$_{2}$ monolayer~\cite{Das2021}, and rescaled it to account for the fluctuational reduction; the latter was deduced from the known experimental data for the bulk compound. Together with the standard formalism for calculating EPC, this forms the basis for addressing superconductivity in monolayer NbSe$_{2}$ from first principles.

\subsection{Role of Charge Density Waves}

The role played by charge density waves in either assisting or opposing superconductivity has been a matter of active debate in the field of unconventional superconductivity. Several recent papers~\cite{Zheng2019,monoCDW} ascribe the notorious overestimation of the superconducting temperature and order parameter to the charge density wave (CDW) effects. We do not believe that CDW alone provides a comprehensive explanation, if at all, for the following reasons:

\vskip 0.2 cm
\begin{itemize}
\item  First of all, overestimation takes place both in the bulk and in single layer calculations of NbSe$_{2}$~\cite{bulkCDW,Sanna,Zheng2019}. Yet in NbSe$_{2}$ suppression of the CDW by pressure or disorder has only a minor effect on the $T_{\rm c}$~\cite{bulkCDW,Hirschfeld}.
\item  NbS$_{2}$ does not exhibit a CDW phase, yet the problem of overestimation there is as severe, if not more so~\cite{Margine1}.
\item  It was shown that in bulk  NbSe$_{2}$ anharmonicity strongly suppresses the tendency to form the CDW~\cite{bulkCDW}, hence it is likely that standard DFT calculations overestimate the CDW amplitude and leads to the partial gapping of the Fermi surface. Bulk calculations for NbSe$_{2}$, accounting for
anharmonicity to suppress CDW at elevated pressure, extrapolate to $T_{\rm c}\approx 12.3$ K and $\lambda\approx1.4$ at zero pressure, a considerable overestimate~\cite{bulkCDW}.
\item Overestimation of the $T_{\rm c}$ was also recently attributed to the empirical treatment of the Coulomb interaction in the Eliashberg formalism compared to the superconducting density functional theory~\cite{Sanna}. Assuming a value of the Coulomb pseudopotential $\mu^*$=0.11 yielded a superconducting $T_{\rm c}$=16~K, whereas a significantly higher value of $\mu^*$=0.28 was necessary to replicate the experimental outcome. Note that, while resorting to an unusually high value of $\mu^*$ reproduces the experimental gap, such Coulomb interactions are not physical even for low density metals, since the value of $\mu^*$ (as opposed to $\mu$) is set by $\log{(E_F/\omega_{ph})}$, and not by the bare Coulomb coupling.
\item  The resistivity in the normal state shows absolutely no detectable feature at the CDW temperature~\cite{bulkCDW2}. If, as suggested in Ref.~[\onlinecite{Zheng2019}], CDW reduces the EPC constant by a factor of seven, the effect on the normal transport would have been dramatic.
\item  In recent experiments~\cite{Ugeda2}, suppressing CDW in single layer NbSe$_{2}$ by disorder (such as Mo doping)  led to $T_{\rm c}$ simultaneously suppressed. 

\end{itemize}

For these reasons, we believe that the effect of CDW on superconductivity
in previous works was overestimated and CDW plays at best a small role in suppressing superconductivity. Instead, in this paper we put emphasis on the pair-breaking effect of magnetic interactions.
\vskip 0.2 in


\section{Results}

\subsection{Theoretical basis}

The recipe for calculating electron-phonon interactions from first principles is well established~\cite{Giustino2017,EPW}. However, the incorporation of the effects of spin-fluctuation warrants reevaluation of the hitherto established protocol. A formalism incorporating spin-fluctuation effects alongside electron-phonon coupling would set the stage to delineate the concomitant landscapes of conventional and  unconventional superconductivity. The momentum-dependent Eliashberg spectral function is given by: 
\begin{equation} 
\afkko\!=\!\NF \sum_{\nu} |\gkkl|^2 \delta(\omega-\oql),
\end{equation}
where $\NF$ is the density of states per spin at the Fermi level, $\gkkl$  are the screened electron-phonon matrix elements, and $\oql$ are the phonon frequencies for a phonon with wavevector $\bq\!=\!\bk\!-\!\bkp$ and branch index $\nu$. 

\begin{figure*}[pt]
	\centering
	\includegraphics[width=\linewidth]{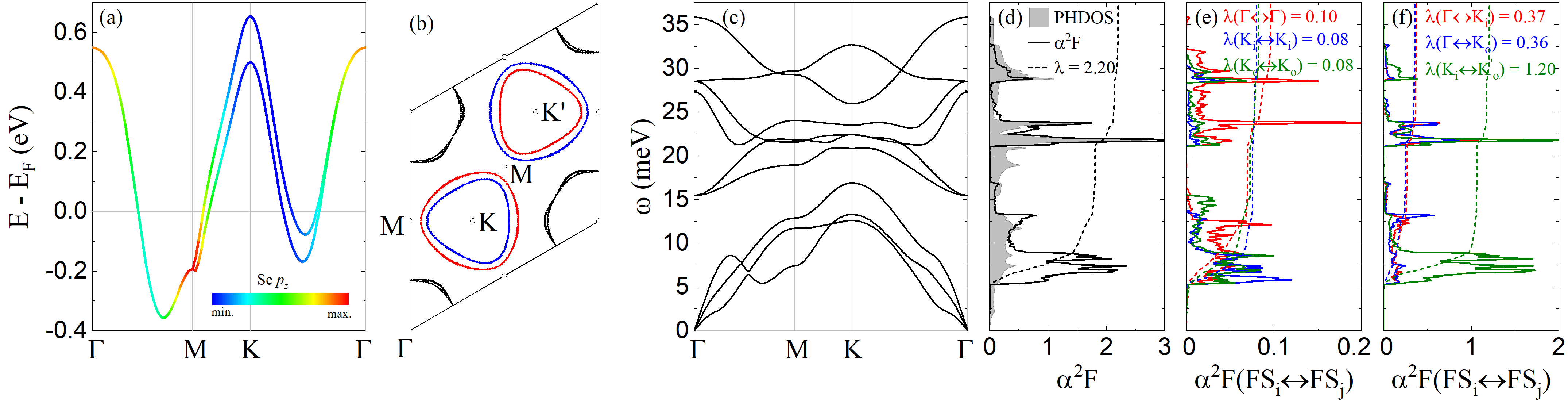}
	\caption{\label{fig1} (a) Calculated band structure with SO coupling in monolayer NbSe$_2$. The color along each band denotes the relative Se $p_z$ character according to the color bar. The orbital character is taken from a calculation without SO coupling. (b) Cross section of the Fermi surface with SO coupling in monolayer NbSe$_2$. Red and blue dots at the $K$ and $K'$ contours denote spin-up and spin-down states. Calculated (c) phonon dispersion, (d) phonon density of states, isotropic Eliashberg spectral function $\afo$ and integrated EPC $\lambda(\omega)$. Decomposition of  $\afo$ and $\lambda(\omega)$ corresponding to (e) intra-pocket $\Gamma-\Gamma$ (red), $K_i-K_i$ plus $K'_i-K'_i$ (blue), and $K_o-K_o$ plus $K'_o-K'_o$ (green) scattering, and (f) inter-pocket $\Gamma-K_i$ plus $\Gamma-K'_i$ (red), $\Gamma-K_o$ plus $\Gamma-K'_o$ (blue), and $K_i-K'_o$ plus $K_o-K'_i$ (green) scattering.}
\end{figure*}

A systematic incorporation of spin fluctuations is less well established, even though the problem goes back to the 1960s \cite{Berk}. The simplest recipe was summarized by D.~Scalapino \cite{Scalapino}, and stipulates
that the effective pairing interaction in the singlet channel
is given by the Eliashberg function $\alpha^2 F_{\rm sf}(\bk,\bkp,\omega)$, defined through the dynamical spin susceptibility $\chi_{\bk\!-\!\bkp} (\omega)$ and (in the modern DFT parlance) the Stoner factor $I$:  
\begin{equation}
\alpha^2 F_{\rm sf}(\bk,\bkp,\omega) = -\frac{3}{2 \pi} \NF I^2 {\rm Im}[ \chi_{\bk\!-\!\bkp}(\omega)].
\label{a2f_sf}
\end{equation}
In the triplet channel the sign is positive (attraction) and the spin-rotation factor 3 is replaced by 1. In practice, the static integrated version of Eq.~(\ref{a2f_sf}), calculated as the Fermi surface average, is universally used:
\begin{equation}
\lambda_{\rm sf} = -\frac{3}{2} \NF\langle I^2 {\rm Re}\chi_{\bf q}\rangle_{\bf q}.
\label{lam_sf}
\end{equation}

More elaborate versions, taking into account ladder diagrams in addition to polarization bubbles, have also been put forward in the following years, most notably by Fay and Appel~\cite{Fay-Appel}, but in proximity to a magnetic instability the only resonant term is the one given by Scalapino~\cite{Scalapino,Scalapino1}. The non-resonant part is usually assumed to be incorporated in the Coulomb pseudopotential.

Equation~(\ref{lam_sf}) has one serious problem however: it completely
neglects retardation effects, implicitly assuming that the characteristic
time scale for the spin fluctuations is the same as for phonons, which
is rarely the case. Because of this, practical applications of this
formalism are plagued by overestimating the SF effect compared to that
of the EPC. For instance, Bekaert {\it et al.}~\cite{Bekaert} recently reported calculations for FeB$_{4}$, and found that Eq.~(\ref{lam_sf}) severely overestimates the effect of SF. To compensate, they have scaled the result by the partial density of the Fe-character states at the Fermi level, even though the original formalism does not provide for that and hybridization effects are supposed to be included in the Stoner factor $I$.

In fact, when a proper frequency dependence is included, the difference in the energy scales between phonons and SF logarithmically reduces the SF induced interaction, pretty much the same way as the Coulomb repulsion is being renormalized to $\mu^*$ \cite{Tol61,Morel}. We include this renormalization implicitly by using Eq.~(\ref{a2f_sf}) instead of Eq.~(\ref{lam_sf}), such that \cite{Kampf1990}
\begin{equation}
\chi_{\bk\!-\!\bkp}(\omega)=\chi_{\bk\!-\!\bkp}(0) P(\omega),
\end{equation}
where
\begin{equation}
P(\omega) = \frac{a \omega}{(\omega-\omega_{\rm sf})^2 + a^2} \theta(\omega_{\rm c} - \omega), 
\end{equation}
with $\omega_{\rm c}=1$~eV the Matsubara frequency cutoff, $\omega_{\rm sf}=0.5$~eV a characteristic frequency for spin fluctuations, and $a=0.1$ a scaling parameter. We estimate the latter two from the calculation of the non-interacting, constant-matrix-element (Lindhard) susceptibility~\cite{Ronald} using the DFT band structure (Fig.~\ref{fig1}(a)), and then further adjust it slightly to match the experimental $T_{\rm c}$.
\vskip 0.2 in
The full formalism now looks as follows:

\begin{eqnarray}
\label{Znorm}  
Z_{\bk}(\oj) = 
   1 + \frac{\pi T}{\NF \oj} \sum_{\bkp j'}  
   \frac{ \ojp \deltakp}{ \sqrt{ \ojp^2+\Delta^2_{\bkp}(\ojp) } } \\ \nonumber
   \times \left[\lambda^{\rm ep}_{\bk,\bkp}(\ojjp) - \lambda^{\rm sf}_{\bk,\bkp}(\ojjp)\right] 
\end{eqnarray} 
\begin{eqnarray} 
\label{Delta}
Z_{\bk}(i\oj) \Delta_{\bk}(i\oj) =  
   \frac{\pi T}{\NF} \sum_{\bkp j'}  
   \frac{ \Delta_{\bkp}(\ojp) \deltakp}{ \sqrt{ \ojp^2+\Delta^2_{\bkp}(i\ojp) } } \\ \nonumber
   \times \left[ \lambda^{\rm ep}_{\bk,\bkp}(\ojjp) + \lambda^{\rm sf}_{\bk,\bkp}(\ojjp) - \mu^*_{\rm c}\right], 
\end{eqnarray} 
This set of coupled nonlinear equations relates the momentum-dependent quasi-particle mass renormalization  function $Z_{\bk}(\oj)$ and superconducting gap function $\Delta_{\bk}(\oj)$.  Here, $\ek$ are the Kohn-Sham eigenvalues, $i\oj=i(2j+1)\pi T$ ($j$ integer) are the fermionic Matsubara frequencies at temperature $T$, and $ \lambda^{\rm ep}_{\bk,\bkp}(\oj)$ and $\lambda^{\rm sf}_{\bk,\bkp}(\oj)$ describe the coupling of electrons to phonons and spin-fluctuations. The two coupling terms can be expressed based on their respective Eliashberg spectral functions:
\begin{equation}
 \lambda^{\rm ep}_{\bk,\bkp}(\ojjp)
= \int_{0}^{\infty} d\omega  \frac{2\omega \afkko}{(\oj - \ojp)^2+\omega^2}, 
 \label{lambda_kko}
\end{equation}
\begin{equation}
\lambda^{\rm sf}_{\bk,\bkp}(\ojjp) =  \int_0^{\infty}d\omega \frac{2\omega \alpha^2 F_{\rm sf}(\bk,\bkp,\omega)}{(\oj-\ojp)^2+\omega^2}. 
\label{lambda_sf}
\end{equation}

A closer look at the expressions~(\ref{Znorm}) and (\ref{Delta}), reveals the fact that the presence of spin fluctuations enhances the quasi-particle mass renormalization by increasing the effective mass of the carriers and suppresses superconductivity in the singlet channel by reducing the effective coupling strength. The credence that this formalism will provide a more fitting description of the experimental superconducting order parameter by establishing electron-phonon coupling and spin fluctuations on an equal footing remains to be ascertained.

\subsection{Computational Results}

Figure~\ref{fig1}(a~-~b) shows the calculated electronic structure of monolayer NbSe$_2$. The Fermi surface consists of three distinct sheets, one centered around the $\Gamma$ point and two around $K$ and $K'$. The broken inversion symmetry in the monolayer leads to the SO interaction splitting each pocket into a pair with spin-up and spin-down states. At the $K$ and $K^{\prime}$ contours, the states with spin-up and spin-down character are depicted as red and blue dots in Fig.~\ref{fig1}(b). Since the splitting near $\Gamma$ is minor, we do not distinguish the states with different spins around this point.

As it has been already pointed out in previous studies~\cite{monoCDW, Zheng2019, Bianco2020}, the lowest-energy branch of the phonon spectra is strongly anharmonic, displaying negative frequencies along the $\Gamma M$ and $M K$ directions. To take care of this unstable mode that drives the system into a CDW transition, we used a larger electronic smearing. With the exception of the soft acoustic mode that hardens and becomes positive, there is no other significant change in the phonon dispersion when the electronic broadening is increased from 0.01 to 0.03~Ry. Our choice of a 0.025~Ry smearing results in a phonon spectrum (see Fig.~\ref{fig1}(c)) which is in good agreement with full anharmonic calculations~\cite{Bianco2020}. 

Based on the topology of the Fermi surface, the Eliashberg spectral function and the EPC strength can be decomposed into intra- and inter-pocket scattering contributions. As shown in Figs.~\ref{fig1}(e)-(f), the inter-pocket scattering is dominant, with more than 50\% of the coupling coming from the inter-pocket scattering between the $K$ and $K^{\prime}$ pockets of the same spin character (i.e., between the states on the inner and outer contours at $K$ and $K^{\prime}$ and vice versa).  In agreement with previous calculations~\cite{Zheng2019}, the superconducting gap is found to be strongly anisotropic and, in the first approximation, can be described as consisting of two gaps (Fig.~\ref{fig2}(a)).  The smaller gap is associated with the $\Gamma$ Fermi sheets, while the larger gap belongs to the $K$ and $K^{\prime}$ sheets. Using $\mu^*_{\rm c}=0.15$, our calculations yield a superconducting critical temperature of 19~K, overestimating even the largest reported experimental value of $\sim$~3.5~K {~\cite{Mak, Sergio}. As discussed, we attribute this discrepancy mainly to the pair-breaking effect of spin fluctuations, and not due to the CDW.

We solve again the anisotropic Eliashberg equations now accounting for spin fluctuations along with the electron-phonon coupling.  The superconducting gap on the Fermi surface at low temperature and the gap distribution as a function of temperature calculated in the presence of SF are displayed in Figs.~\ref{fig2}(b) and (d). Under the influence of spin fluctuations, the two-gap structure is maintained, but the superconducting gap and the corresponding critical temperature are drastically reduced. Using a spin-fluctuation frequency $\omega_{\rm sf}=0.5$~eV, $a=0.1$, and $\mu^*_{\rm c}=0.15$, we get $T_{\rm c}=3.2$~K in good agreement with the experimental values. 

\begin{figure}[pt]
	\centering
	\includegraphics[width=1.0\linewidth]{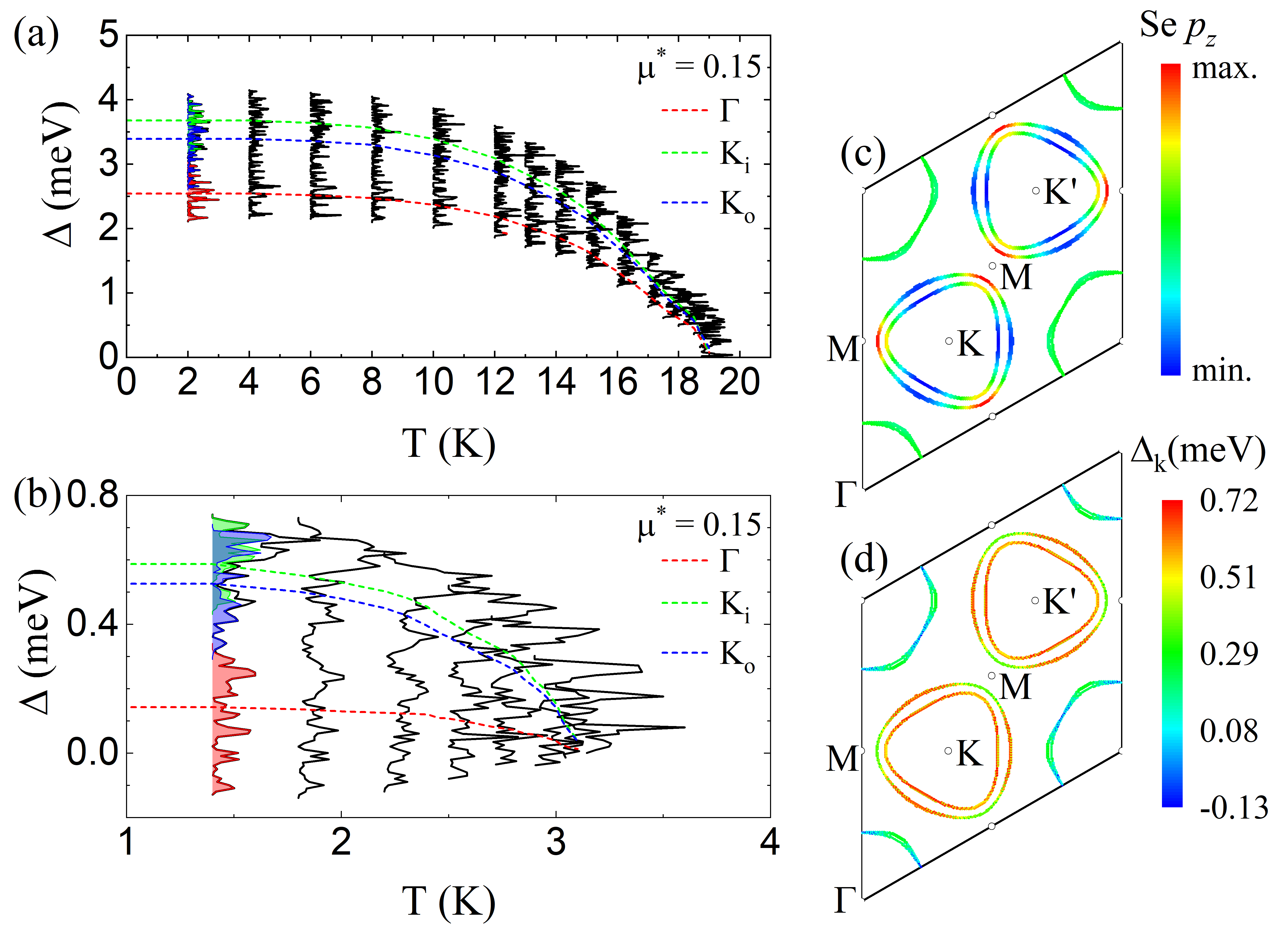}
	\caption{\label{fig2} Superconducting gap distribution as a function of temperature obtained by solving the anisotropic Eliashberg equations~(\ref{Znorm}) and (\ref{Delta}) (a) without and (b) with the inclusion of spin fluctuations. Red, green and blue shaded area  represent contributions to the superconducting gap associated with the FS sheets at $\Gamma$, $K_i$, and $K_o$, while dashed red, green, and blue curves are averages of the anisotropic solutions. (c) Cross section of the Fermi surface with Se $p_z$ character. (d) Superconducting gap $\Delta_{\bk}$ at 1.4~K on the Fermi surface corresponding to anisotropic Eliashberg calculations with the inclusion of spin fluctuations.}
\end{figure}

\section{Discussion}

\begin{table}[ptb]%
\centering
\caption{Calculated electron-phonon coupling matrix $\lambda_{ij}$. The subscripts ``o'' and ``i'' stand for the outer and inner pocket around the corresponding point. We do not distinguish the outer and inner pockets around $\Gamma.$ The first line gives the partial DOS on each Fermi surface pocket in states/(eV f.u.). The  last two lines give the two largest eigenvalues of the $\lambda_{ij}$ matrix, and the eigenvectors giving the relative order parameters.}

\begin{tabular}
[c]{llllll}
&$\Gamma$ & $K_i$&$K_o$&$K'_i$&$K'_o$ \\
\hline
$N(E_F)$ & 0.839 & 0.316 & 0.367 & 0.316 & 0.367 \\
\hline

$\Gamma$ & 0.126 & 0.123 & 0.118 & 0.123 & 0.118\\
$K_i$ & 0.327 & 0.140 & 0.000 & 0.000 & 1.051\\
$K_o$ & 0.270 & 0.000 & 0.123 & 0.905 & 0.000\\
$K'_i$ & 0.327 & 0.000 & 1.051 & 0.140 & 0.000\\
$K'_o$ & 0.270 & 0.905 & 0.000 & 0.000 & 0.123\\
\hline
1.24 & -0.21 & -0.51 & -0.47 & -0.51 & -0.47\\
\hline
1.1 & 0.00 & -0.52 &0.48 &0.52 & -0.48\\
\end{tabular}%
\label{EPCtable}%
\end{table}

\subsection{Superconductivity and symmetry of pairing function}

Let us first start with the results of the EPC calculation only, as these already uncover unexpected and important physics. The first observation, as mentioned, is that unmitigated EPC is way too strong to be consistent with the experiment, calling for spin fluctuations. Regardless of this, the calculated EPC is strongly nonuniform.
The EPC is strongly dominated by the
$K_i-K'_o$ and the equal by symmetry $K_o-K'_i$ coupling (note that $\lambda_{ij}
=V_{ij}N_j$, where $V$ is a symmetrical matrix and $N_j$ is partial DOS). This implies that the order parameter will be similar in magnitude on the $K$ sheets, but the phase between the $K_i-K'_o$ and $K_o-K'_i$ manifolds may be varied without a big
loss of the pairing energy. The order parameter on the $\Gamma$ pockets 
will be mostly induced by the interband proximity effect and is expected to be relatively small. All this is corroborated by our full Eliashberg calculations. 

Before analyzing the pairing symmetry, we shall make an important note. In regular, non-Ising superconductors ($i.e.$, not spin-orbit split, but possibly SO-influenced)  a standard way to analyze the pairing symmetry, whether on the level of the simple linearized BCS equations, or full anisotropic Eliashberg calculations, is to assign a complex value of the order parameter to each point on each Fermi surface, and proceed from there. The standard signature of a triplet pairing is the phase shift of $\pi$ ($i.e.$, a sign changed between the ${\bf k}$ and  ${\bf -k}$ points).

Importantly, this is not a unique procedure and depends upon the choice of the phase gauge in the normal state between different  ${\bf k}$-points, which sometimes leads to nontrivial ramifications \cite{Over}. It becomes even more nontrivial in case of an Ising superconductor. To illustrate this, we will use as order parameters anomalous averages as defined in Ref.~[\onlinecite{Darshana}]:
\begin{equation}%
\begin{split}
d_{o,\mathbf{k}}  &  =\overline{\left\vert K,o,\uparrow\right\rangle
\left\vert K^{\prime},o,\downarrow\right\rangle }\\
d_{i,\mathbf{k}}  &  =-\overline{\left\vert K,i,\downarrow\right\rangle
\left\vert K^{\prime},i,\uparrow\right\rangle }%
\end{split}
\label{eq:d0,di}%
\end{equation}
Note that, as opposed to a regular, Kramers-degenerate superconductor, there is no
such thing as $d_{\mathbf{-k}} $, because a $K,o$ state has $only$ the $\uparrow$, and a $K,i$ $only$ the $\downarrow$ one. We illustrate this in Fig.~\ref{sf}: while a {\it nonrelativistic} bilayer has topologically the same $K$-like Fermi surface, also splits around $K$ and $K'$ points, it actually has separate order parameters for the  ${\bf k}$ and ${\bf -k}$ points and thus two options depicted in  Fig.~\ref{sf}(a,b). The former corresponds to s-wave, and the latter to the f-wave pairing. In an unlikely case that the EPC in a bilayer is dominated by the $K_o-K'_i$ scattering, both states are close in energy, despite the interaction being purely EPC. 

In our case of an Ising superconductor the dominance of the EPC $K_o-K'_i$ scattering 
fixes the phases as shown in Fig. \ref{sf}(c). This essentially excludes the possibility
of a predominantly triplet state, albeit generates a small ($\approx 10$\%) triplet
admixture. 

Let us now turn to spin fluctuations (Table II). Quantitatively, as conjectured in Ref.~[\onlinecite{Darshana}], the SF coupling in the $K-K'$ channels is small, and so is the intraband coupling. There is a sizeable $K_o-K_i$ coupling, which does favor
triplet, but it is not strong enough, and, surprisingly, an even stronger contribution appears in the $K-\Gamma$ channels. As a result, not only a triplet f state, corresponding to  Fig. \ref{sf}(d), is unstable compared to the s state, it is not even competitive, but an s$_\pm$ state, where the order parameter in the $\Gamma$ pocket is flipped compared to the $K$ pockets is competitive. At the level of accuracy 
available in our calculations, we can exclude the s-f Leggett mode, recently proposed for tunneling measurements~\cite{Ugeda2}, but  cannot exclude the possibility 
of a Leggett mode associated with the phase fluctuations between the $\Gamma$ and
$K$ pockets. This we will address in more detail later. We can also exclude the recently proposed nematic superconductivity
ascribed to a close competition between s-wave and a higher-angular-momenta
state~\cite{Jorg}.

It is still instructive to compare our results with the simple linearized BCS solution that requires the order parameters near the transition temperature to be proportional to the eigenvectors of the matrix $\lambda$, corresponding to its largest eigenvalue. Diagonalizing the matrix in Table~\ref{EPCtable}, we get the largest eigenvalue $\lambda_{\max} =1.24$, and the corresponding order parameters as shown in the penultimate line in the same table. 

\begin{figure}[tp]
	\centering
	\includegraphics[width=.46\linewidth]{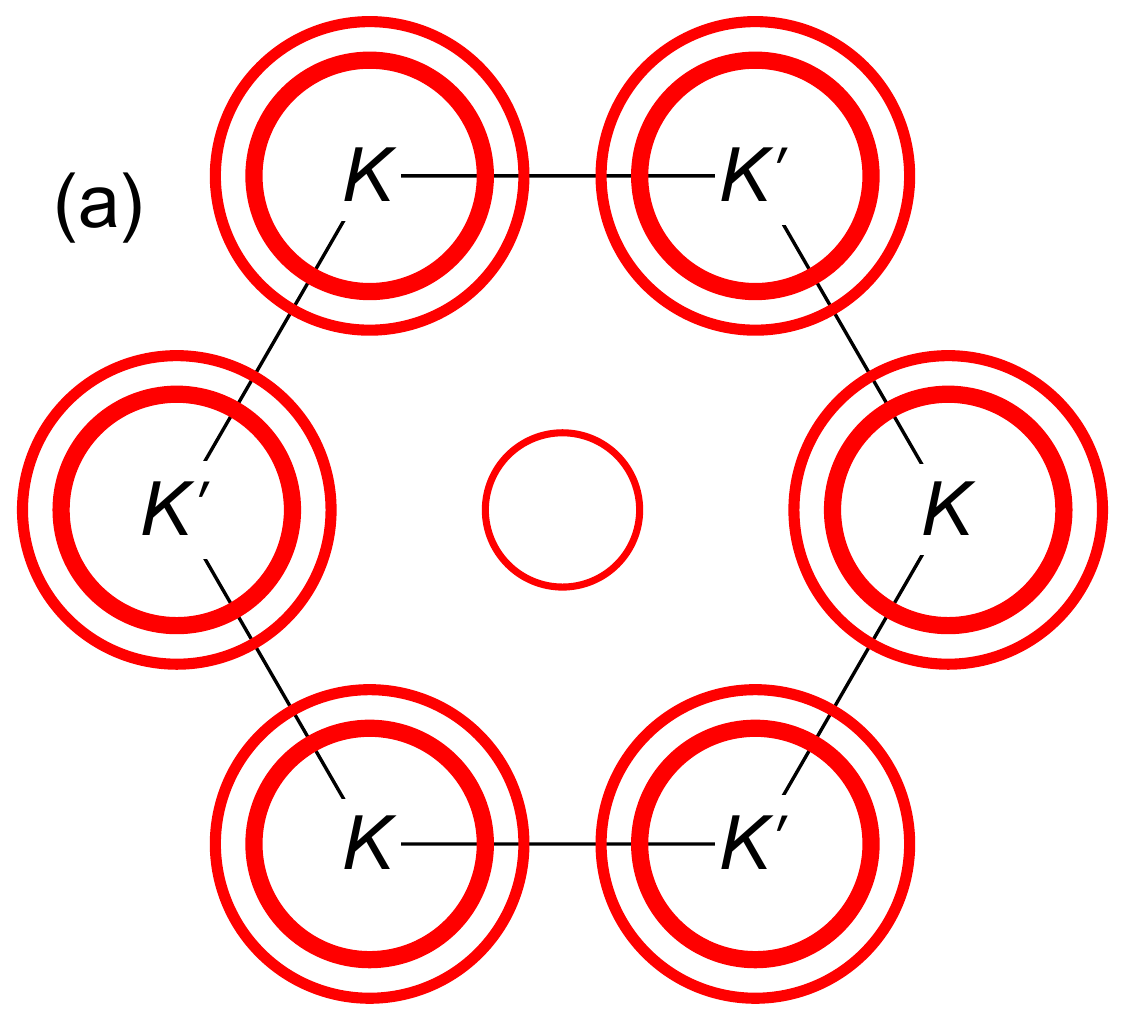}	\includegraphics[width=.46\linewidth]{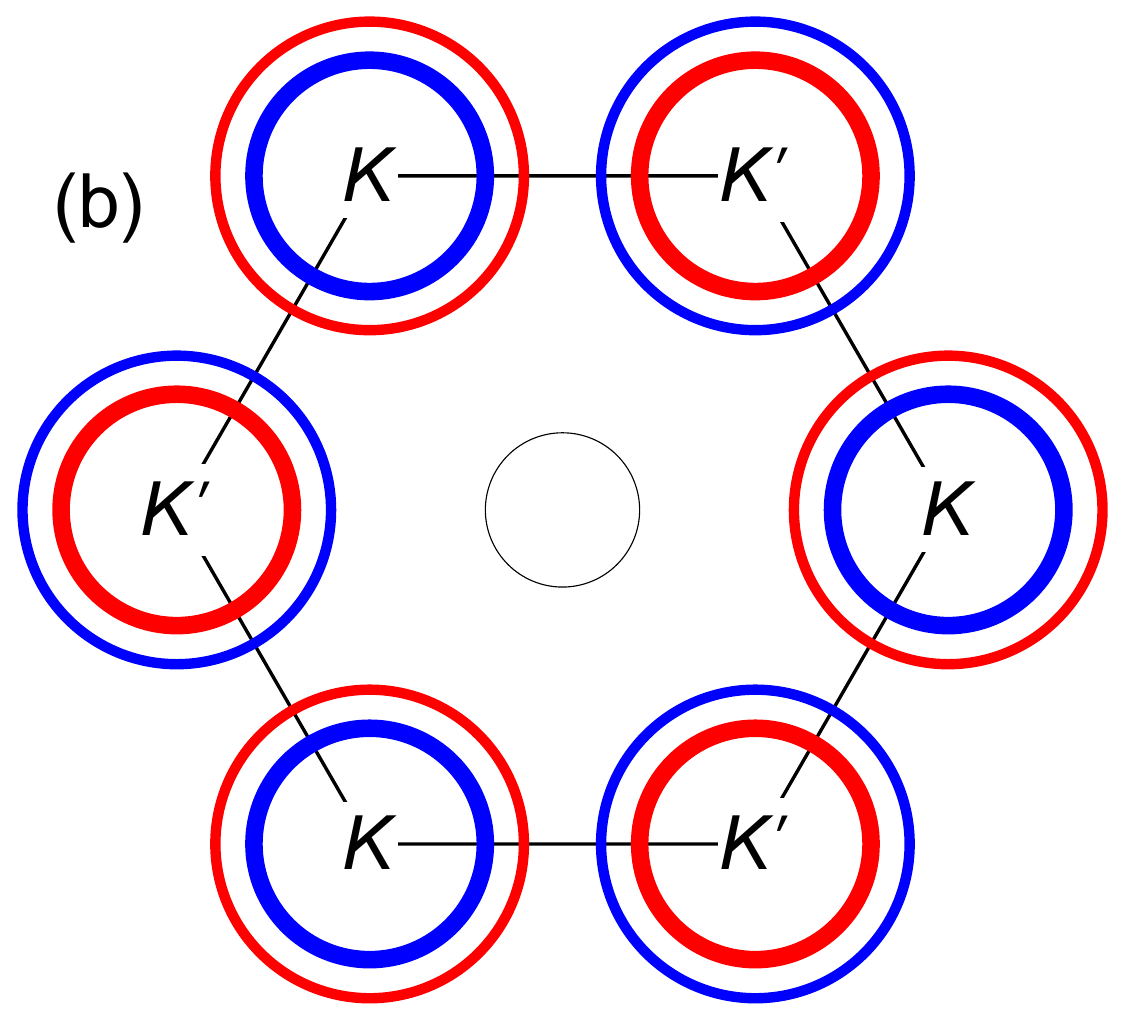}
	\includegraphics[width=.46\linewidth]{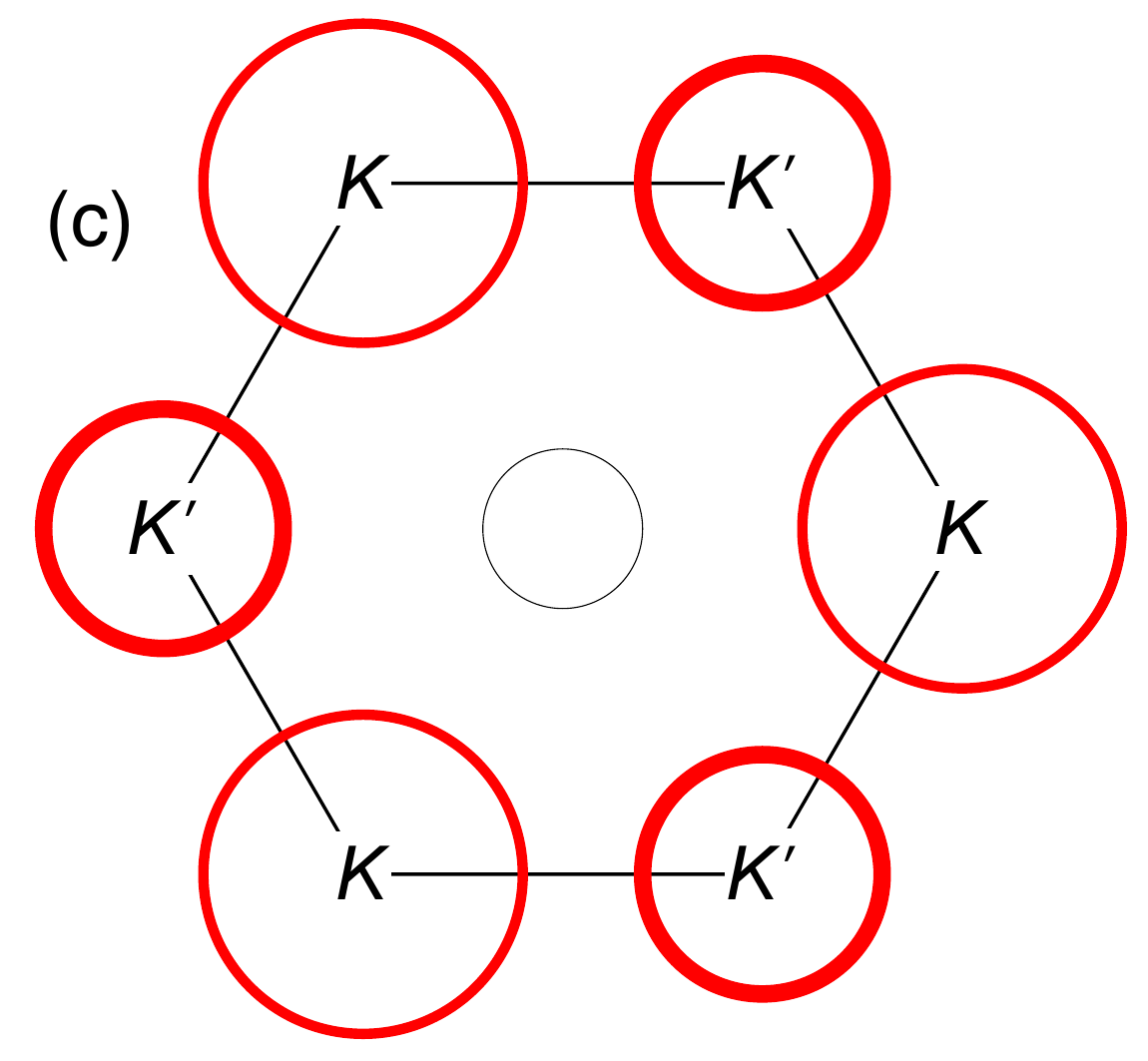}	\includegraphics[width=.46\linewidth]{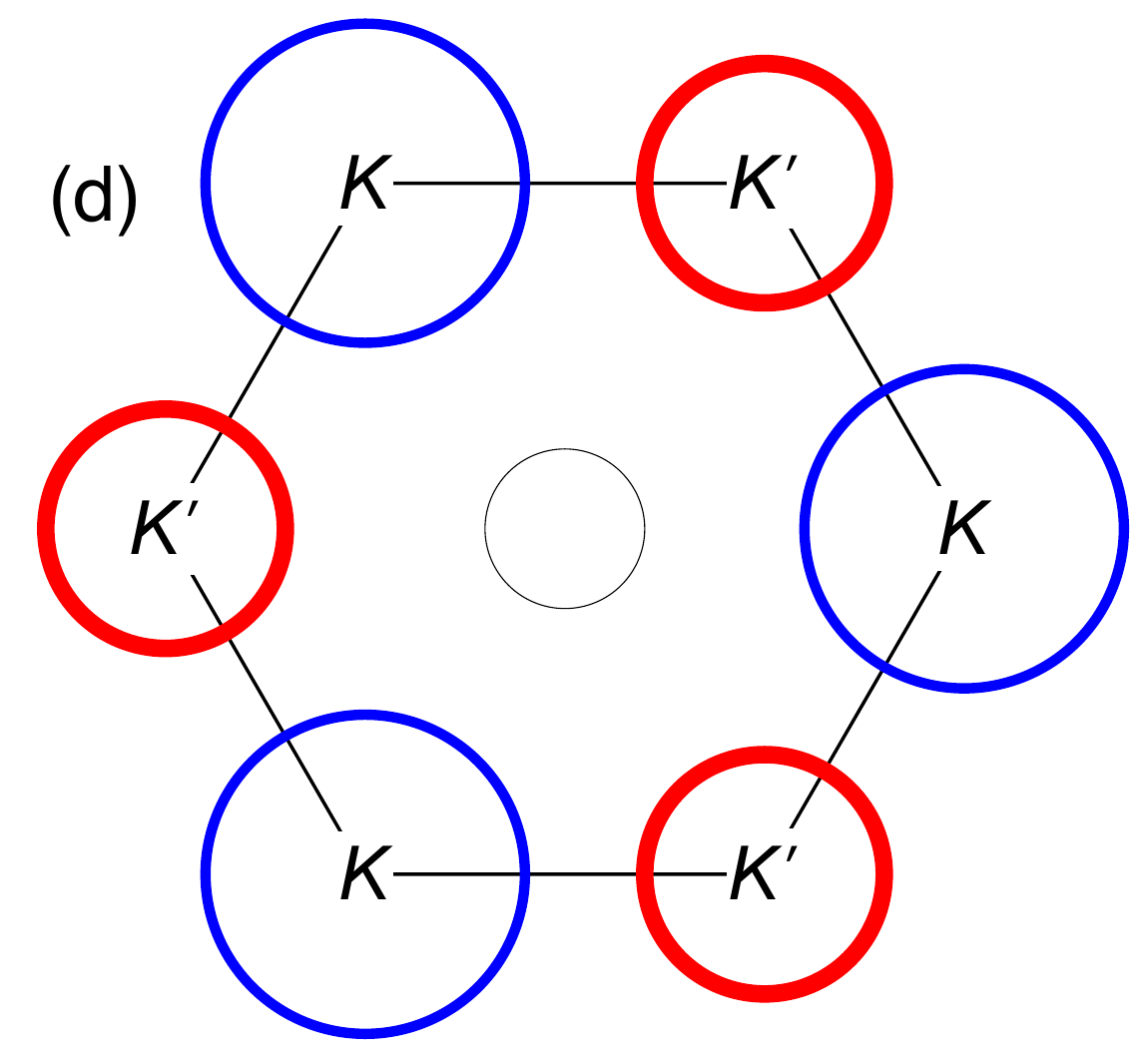}
	\caption{Schematic illustration of possible pairing symmetry in a regular
	superonductor with hybridization-split, Kramers-degenerate Fermi surfaces,
	and in an Ising superconductor. Splitting and superconductivity
	on the $\Gamma$ pocket are neglected. The color shows the sign of the order
	parameter. Panels (a) and (c) correspond to an s-wave, and (b) and (d) to a 
	f-wave symmetry. Note that in the IS the order parameter cannot be independently defined for ${\bf k}$ and ${\bf -k}$, so only one is shown.
	}\label{sf} 
\end{figure}

There is rather little difference between the inner and the outer pockets,
again in agreement with the full Eliashberg solution, about 10\%. The $\Gamma$
pocket order parameter is less than half of those on the  $K$ pockets. Earlier we have discussed the   potential ramifications of this for tunneling.

Amazingly, and rather unexpectedly, the second largest eigenvalue is 1.1 only 12\% smaller. In the weak coupling regime, this corresponds to an incipient superconducting state with a transition temperature that is only moderately 
smaller than the one for the leading instability. So, the leading instability is, basically, a two-gap s-wave superconductivity, not dissimilar to that in MgB$_2$.



\begin{table}[ptb]%
\centering
\caption{Same as in Table~\protect\ref{EPCtable}, but for 
spin-fluctuations induced coupling.}

\begin{tabular}
[c]{llllll}
&$\Gamma$ & $K_i$&$K_o$&$K'_i$&$K'_o$ \\
\hline
$N(E_F)$ & 0.839 & 0.316 & 0.367 & 0.316 & 0.367 \\
\hline
$\Gamma$ & 0.543 & 0.132 & 0.182 & 0.133 & 0.182\\
$K_i$ & 0.353 & 0.087 & 0.222 & 0.095 & 0.053\\
$K_o$ & 0.416 & 0.191 & 0.101 & 0.045 & 0.125\\
$K'_i$ & 0.353 & 0.095 & 0.053 & 0.087 & 0.222\\
$K'_o$ & 0.416 & 0.045 & 0.125 & 0.191 & 0.101\\
\hline
1.00& -0.57&-0.39&-0.43&-0.39&-0.43\\
\hline
-0.17&0.0&0.505&-0.5&-0.506&0.493\\
\end{tabular}%
\label{SFtable}%
\end{table}

\subsection{Implications on possible Leggett mode}


Recent tunneling data have suggested the appearance of a superconducting collective mode interpreted as a Leggett mode between the s-wave state and a proximate f-wave triplet channel \cite{Ugeda2}. In this section,  we use our first principles results to examine the possibility that this mode is due to fluctuations of the order parameter phase between the $K$ pockets and the $\Gamma$ pockets.

In this section, we have delineated an analytical evaluation of a self-consistent solution of the Bardeen-Cooper-Schrieffer theory of superconductivity, based on parameters derived from first principles calculations. 
In order to solve for the Leggett modes, we adopt the following scheme as described below. 

1. First, we assume that the superconducting order parameter $\Delta$ varies very little within each individual sheet of the NbSe$_{2}$ Fermi surface (FS), while differing significantly between the different FS sheets. Such  assumptions lead to the following expression for  the superconducting order parameter:
\begin{equation}
\Delta _{i}=\sum_{i}\Lambda_{_{ij}}\Delta _{j}F(\Delta _{j},T)
\end{equation}
where
\begin{equation}
F=\int_{0}^{\omega_{D}}dE \: \tanh\:((\sqrt{E^{2}+\Delta^{2}})/2T)/\sqrt{E^{2}+\Delta^{2}}
\end{equation}
Note that here the matrix $\Lambda_{_{ij}}$ characterizes the electron-phonon interaction and can be expressed in terms of the convolution of the pairing interaction due to electron-phonon coupling and the band-resolved density of states as $\Lambda_{_{ij}}= V_{ij}N_{j}$, whereas the temperature dependence of the superconducting gap function is dictated by the functional form $F$.

2. Second, in order to incorporate the effect of spin-fluctuation contributions to the superconducting pairing interaction in the monolayer, a fluctuation parameter $\alpha$ has been defined to establish the effect of spin-fluctuations on the same footing as that of electron-phonon coupling. The material specific fluctuation term $\alpha$, that accounts for the effect of renormalized spin-fluctuation is of negative sign. We define the renormalized  matrix $\Lambda_{ij} = \Lambda^{ep}_{ij}+\alpha  \Lambda^{sf}_{ij}$, which denotes the total (electron-phonon coupling and spin fluctuation) pairing interaction term for solving the BCS equations.

3. Third, we choose starting values of the gap parameters $\Delta_{1}$ and $\Delta_{2}$, respectively, where $\Delta_{1}$ and $\Delta_{2}$ refer to the superconducting order parameters for the $\Gamma$ and $K$ points of the FS sheets and solve the above BCS equations self-consistently in order to obtain an analytical solution of the superconducting gap equation as a function of temperature. 

4. Fourth, we determine the value of the material-specific spin-fluctuation parameter $\alpha$, and hence the renormalized matrix $\Lambda_{_{ij}}$,  heuristically by incorporating the values of $\Delta_{1}$ and  $\Delta_{2}$ obtained from our first principles calculations of electron-phonon coupling and spin fluctuations obtained from the solution of the Eliashberg equations from first-principles using the EPW software. 

5. Finally, we utilize the parameters obtained from first-principles and the analytical solution of the self-consistent BCS equations, in order to provide a quantitative estimate of the frequency of the superconducting Leggett mode for comparison with experimental observations. 
This is achieved by solving the following equation
\begin{equation}
\omega_{L}^{2}=\frac{4\Delta_{1}\Delta_{2}V_{12}}{det\: V}\; \frac{N_{1}f_{1}+N_{2}f_{2}}{N_{1}N_{2}f_{1}f_{2}},
\end{equation}
where $f(\omega)={\sininv \omega}/{\omega \sqrt{1-\omega^{2}}}$. Here $\Delta_{1}$ and $\Delta_{2}$ again refer to the superconducting order parameters for the $\Gamma$ and $K$ points respectively of the Fermi surface. Thus,  self-consistent solution of the above set of equations lead to the quantitative estimate of the Leggett frequency.
\vskip 0.5 cm
\begin{figure}[pt]
	\centering
	\includegraphics[width=0.9\linewidth,height=0.8\linewidth]{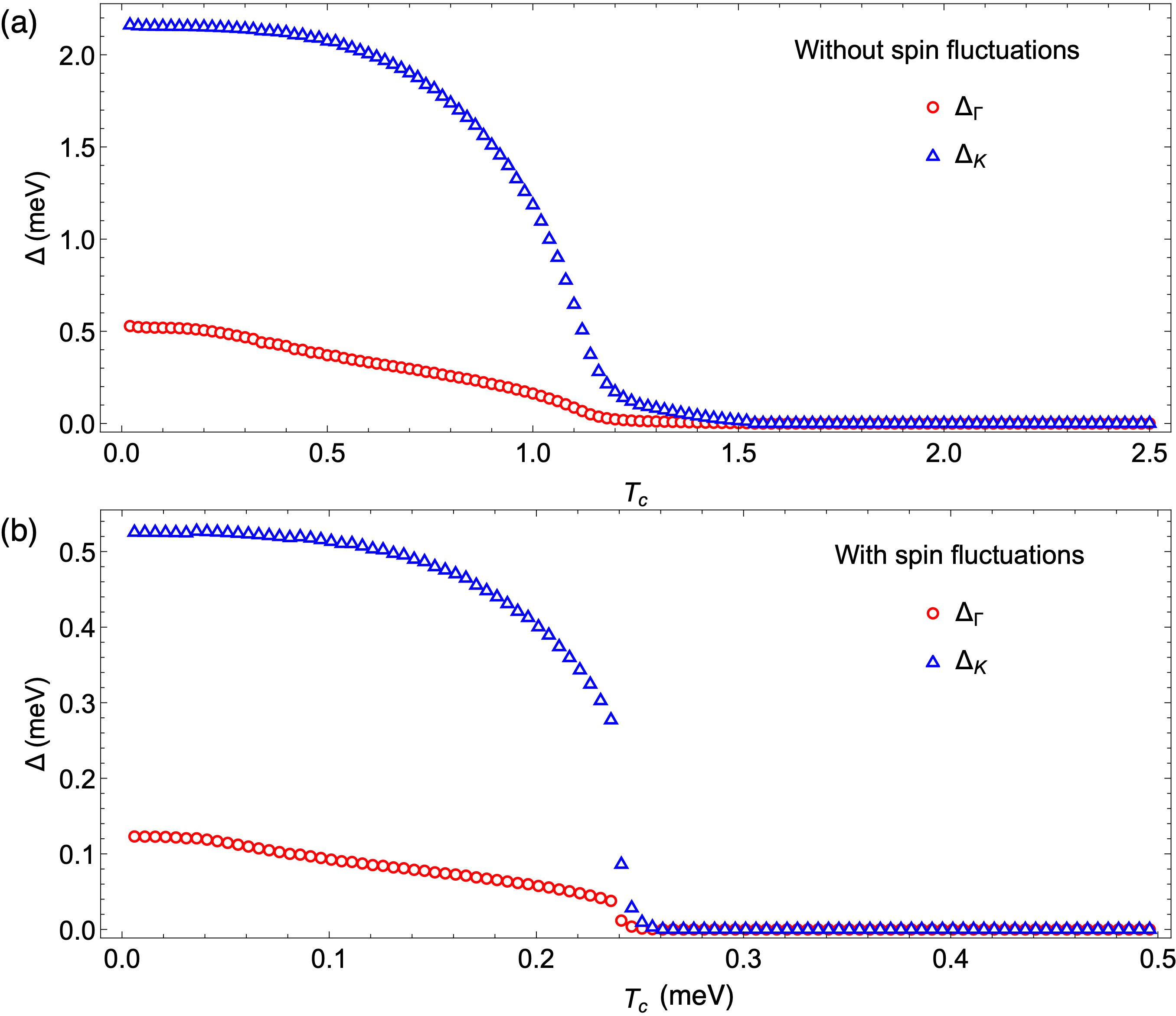}
	\caption{\label{fig5} Analytical evaluation of the pairing interaction, by fitting the two-gap BCS equations to that of our first principles calculations including (a) electron-phonon interaction and (b) both electron-phonon interaction and spin fluctuation effects evaluated utilizing density functional theory artificially stabilized by Hubbard interactions combined with spin-spiral methodology.}
\end{figure}

In order to further reconstruct the gap structure from first-principles calculations, we proceed with the pairing interaction matrix derived from the solution of the Eliashberg equations, and implement required modifications to analytically solve the BCS equations. Figs. 4(a) and (b) illustrate the analytical solution of the pairing interaction, by fitting the two-gap BCS equations with that of our first principles calculations without and with the inclusion of spin fluctation effects respectively, in addition to electron-phonon coupling. The ensuing elements of the interaction matrix utilized for our analytical solution of the self-consistent BCS equations, are $V_{11}$ = 0.2957, $V_{12}$ = $V_{21}$ = 0.0558, and $V_{22}$ = 0.5997 eV, respectively. The density of states estimated from first-principles $N_{1}$ and $N_{2}$ are 0.8392 and 1.365 states/(eV f.u.), which correspond to that of the $\Gamma$ and $K$ segments of the Fermi surface. Finally, the magnitudes of the superconducting gaps resulting from solving the Eliashberg equations $\Delta_{1}$ and $\Delta_{2}$ are 0.15 and 0.55 meV for the $\Gamma$ and $K$ segments of the Fermi segments. We note that, the frequency of the Leggett mode obtained from resonant phase fluctuations, should, in principle, be larger than the magnitude of the minimum superconducting gap of 0.15 meV from that of the $\Gamma$ point.
Interestingly, our calculation of the frequency of the Leggett mode, yields a value of 0.29 meV for the Leggett mode frequency.

 In the tunneling experiments, that recently reported the observation of soft collective modes in monolayer NbSe$_{2}$~\cite{Ugeda2}, new satellite peaks are seen at integer multiples of the fundamental frequency in the single layer samples. These satellites are thought to have their origin in collective Leggett modes in the experimentally grown monolayers. Experimental conductance spectrum revealed an average BCS gap of $\Delta_{BCS}$ = 0.4 meV, a resonant Leggett mode frequency $\Omega_{L}$ = 0.53 meV and the ratio $\Omega_{L}/2\Delta_{BCS}$ = 0.66. Our self-consistent analytical solution and first-principles calculations including electron-phonon coupling and spin-fluctuation contribution correspond to an estimated average $\Delta_{BCS}$ = 0.35 meV, the frequency of the collective Leggett mode $\omega_{L}$ = 0.29 meV and the ratio $\omega_{L}/2\Delta_{BCS}$ = 0.41. A final note worth mentioning in passing is that while an amplitude Higgs mode~\cite{Higgs, Higgs1} has also been observed in the superconducting bulk NbSe$_{2}$~\cite{NbSe2-Higgs1,NbSe2-Higgs2} due to the mixing with collective CDWs, such a mode can be discarded in monolayer NbSe$_{2}$, since the CDW mode energy is considerably larger than the superconducting order parameter in the monolayer case.
\vspace{0.2 in}
\section{Conclusion}
Utilizing state of the art DFT and Wannier interpolation formalism, we have calculated the momentum-resolved electron-phonon coupling interaction in the single layer NbSe$_2$. The two main findings are: (1) the overall strength of this interaction is such that the superconducting critical temperature and the gap parameter are substantially overestimated, and (2) the leading contribution to the electron-phonon coupling comes from the intepocket scattering between the K and K' pockets of the Fermi surface.
We suggest that the phonon-induced superconductivity in single NbSe$_2$,
and likely in other similar materials, bulk or monolayer, is weakened by spin-fluctuations. 

We find that, if the standard static formulation of the Berk-Schrieffer-Scalapino formalism is adopted, realistic estimates of the strength of electron-spin-fluctuation coupling lead to complete suppression of superconductivity, in dramatic contradiction with the experiment. We argue that this is due to neglect of the retardation effects, weakening the effect of spin-fluctuations logarithmically, analogous to, but not as strong as the famous Tolmachev-Morel-Anderson renormalization of the Coulomb repulsion.

If the latter is accounted for, the structure of the superconducting order parameter changes notably from a pure phonon mechanism, albeit the fact that the strongest interaction occurs in the K-K' channel, remains, and even becomes stronger, and the overall agreement with the experiment is satisfactory. We find that the leading instability is in the s$_{++}$ wave channel, and the 
subleading one in the s$_\pm$ channel, where the sign of the (smaller) order parameter on the $\Gamma$-centered pocket is flipped with respect to the K, K' pockets.

We estimate the frequency of the Leggett mode driven by this subleading instability, and find it to be in reasonable agreement with the recently claimed Leggett mode observation in the STM spectroscopy.

\vspace{0.2 in}

\section{Computational details} 
\noindent First-principles calculations were performed with density functional theory (DFT) using the Quantum ESPRESSO (QE)~\cite{QE2017} code. We employed optimized norm-conserving Vanderbilt (ONCV) pseudopotentials~\cite{Hamann2013,Schlipf2015} with the Perdew-Burke-Ernzerhof (PBE) exchange-correlation functional in the generalized gradient approximation~\cite{GGA}, where the Nb $4s^2  4p^6 4d^3 5s^2$ and Se $4s^2 4p^4$ orbitals were included as valence electrons. All calculations were performed for the experimental lattice parameters at ambient pressure~\cite{Weber2011} with relaxed internal coordinates. We used a plane wave kinetic-energy cutoff value of 80 Ry, and the electronic and vibrational Brillouin zones (BZ) were sampled using $24\times 24 \times 1$ and $12 \times 12 \times 1$ points, respectively. A Methfessel and Paxton smearing~\cite{MP1989} width of 0.025~Ry was used in order to resolve the CDW instability. 

The  superconductivity calculations were performed with a modified version of the EPW code \cite{Giustino2007,EPW, Margine2013}. We used 22 maximally localized Wannier functions~\cite{WANN1,WANN2} (five $d$-orbitals for each Nb atom and three $p$-orbitals for each Se atom)  and a uniform $\Gamma$-centered $12 \times 12 \times 1$ electron-momentum grid. Eqs.~(\ref{Znorm})-(\ref{Delta}) were evaluated on a uniform $240 \times 240 \times 1$~\textbf{k}-point grid and a uniform $120 \times 120 \times 1$ \textbf{q}-point grid. The Dirac deltas were replaced by Gaussians of width 2.5~meV (electrons) and 0.1~meV (phonons), and the Matsubara frequency cutoff was set to 1~eV.

The momentum-dependent static spin susceptibility $\chi_{\bq}(0)$ in the random-phase approximation (RPA) and the Stoner parameter $I$ were obtained in our previous work~\cite{Das2021}. The DFT Stoner factor was found to be $I=0.646$ eV/f.u. and the spin susceptibility was divided by $6.466 \times 10^{-5}$ to convert from emu/mol to 1/eV units used in the current study. The analytical solution of the BCS equations were performed utilizing the Mathematica, Wolfram Language software package.~\cite{Mathematica}

\noindent The work at GMU (S.D. and I.I.M.) was supported  by ONR through grant N00014-20-1-2345.
H.P. and E.R.M. acknowledge support from the National Science Foundation (Award No. 2035518). D.F.A. was supported  by the U.S. Department of Energy, Office of Basic Energy
Sciences, Division of Materials Sciences and Engineering,
under Award No. DE-SC0021971. This work used the Extreme Science and Engineering Discovery Environment (XSEDE)~\cite{XSEDE} which is supported by National Science Foundation grant number ACI-1548562.  Specifically, this work used Comet at the San Diego Supercomputer Center through allocation TG--DMR180071. 

\bibliography{spin2.bib}
\end{document}